\documentclass[]{interact}

\usepackage{epstopdf}
\usepackage[caption=false]{subfig}

\usepackage[numbers,sort&compress]{natbib}
\bibpunct[, ]{[}{]}{,}{n}{,}{,}
\renewcommand\bibfont{\fontsize{10}{12}\selectfont}
\theoremstyle{plain}

\theoremstyle{definition}

\theoremstyle{remark}

\begin{document}


\title{Parameter Estimation for Grouped Data Using EM and MCEM Algorithms }
\author{
\name{Zahra A. Shirazi\textsuperscript{a}, Jo\~ao Pedro A. R. da Silva\textsuperscript{b} and Camila P. E. de Souza\textsuperscript{a}}
\affil{\textsuperscript{a} The University of Western Ontario, ON, Canada; \textsuperscript{b} University of S\~ao Paulo, SP, Brazil}
}


\maketitle

\begin{abstract}
Nowadays, the confidentiality of data and information is of great importance for many companies and organizations. For this reason, they may prefer not to release exact data, but instead to grant researchers access to approximate data. For example, rather than providing the exact measurements of their clients, they may only provide researchers with grouped data, that is, the number of clients falling in each of a set of non-overlapping measurement intervals. The challenge is to estimate the mean and variance structure of the hidden ungrouped data based on the observed grouped data. To tackle this problem, this work considers the exact observed data likelihood and applies the Expectation-Maximization (EM) and Monte-Carlo EM (MCEM) algorithms for cases where the hidden data follow a univariate, bivariate, or multivariate normal distribution. Simulation studies are conducted to evaluate the performance of the proposed EM and MCEM algorithms. The well-known Galton data set is considered as an application example. 
\end{abstract}

\maketitle

\newpage
\section{Introduction}
Nowadays, protecting data confidentiality, security, and integrity is of great importance for governments, organizations, and companies \cite{Chen2018,Huang2016,Minoiu2009,Wua2007}. For these reasons, these institutions might not release exact raw data to researchers, analysts, or even the public. Rather, they prefer to release data such as household income, house prices, insurance losses, profits, and age in an interval format.
The interval format can contain either grouped data \cite{Velez2015} or symbolic data \cite{SymbolicData}. This work focuses on grouped data, where for a particular variable, only the intervals and the frequency of observations falling into each interval are known. Table \ref{table:Univariate Grouped data} shows how univariate grouped data can be represented.

\begin{table}[h]
\begin{center}
\begin{tabular}{c|c}
Interval& Frequencies \\ \hline
$[a_0,a_1)$ & $n_1$ \\
$[a_1,a_2)$ & $n_2$\\
\vdots& \vdots \\
$[a_{k-1},a_k)$ & $n_{k}$\\ \hline
Total & $n$
\end{tabular}
\label{table:Univariate Grouped data}
\end{center}
\caption{Univariate grouped data representation.} 

\end{table}

As can be seen from the grouped data representation in Table \ref{table:Univariate Grouped data}, these data are histogram-based and, therefore, continuous. Continuous data can follow different distributions, including normal, log-normal, and Weibull. 
Many studies have been conducted on grouped data from different perspectives. Tallis \cite{Tallis1967} has obtained approximate maximum likelihood estimates of the parameters for univariate and multivariate grouped data. Stewart \cite{Stewart1983} has dealt with the problem of estimating the parameters of a linear model using data in which the dependent variable is only observed to fall in certain intervals on a continuous scale, with its actual values remaining unobserved. Mclachlan and Jones \cite{McLachlan1988} have considered the fitting of finite mixture models to univariate grouped and truncated data using the Expectation-Maximization (EM) algorithm \cite{Dempster1977,Mclachlan2007}. Heitjan \cite{Heitjan1989} has considered Bayesian methods to analyze this type of data. In another study, Heitjan \cite{Heitjan1991} has applied Newton-Raphson's method and the EM algorithm to find parameter estimates of bivariate regression analysis for grouped data. Cadez et al. in \cite{cadez2002maximum} have extended the work in \cite{McLachlan1988} to multivariate grouped data by using numerical techniques to evaluate the multidimensional integrals at each iteration of the EM algorithm. Wengrzik and Timm \cite{Wengrzik2011} have studied the performance of different methods for fitting a two-component Gaussian mixture model to univariate grouped data. Velez and Correa \cite{Velez2015} have estimated the mean, variance, and coefficient of variation for univariate grouped data using their proposed bootstrap method. More recently, Teimouri in \cite{SkewNormal} has applied the EM algorithm on univariate grouped data arising from a mixture of skew-normal distributions. 
 
 

The aim of this study is to find the parameter estimates for grouped data when they are normally distributed for the univariate, bivariate, and multivariate cases using the exact form of the likelihood. Therefore, the estimation approach of \cite{McLachlan1988} and \cite{Mclachlan2007} for univariate grouped data with missing counts is considered  and extended to the univariate, bivariate, and multivariate cases without missing counts using both the EM and MCEM algorithms.
To the authors' knowledge, no other study has yet presented the exact formulae of EM parameter estimates for the bivariate and multivariate normal grouped data, as is done in this work. This work also contains the formulae to obtain standard errors for the EM and MCEM mean estimates. 
In summary, three possible approaches for parameter estimation of grouped data are presented: 
1) maximum likelihood estimation (MLE) by numerical optimization of the exact grouped data likelihood (Exact MLE), 2) maximizing the exact likelihood using the EM algorithm, and 3) same as (2), but using the MCEM algorithm. All three methods are implemented in R and available at https://github.com/desouzalab/infgrouped.

This study is organised as follows. In Section 2, the estimation methods for grouped data are presented. In Section 2.1, univariate normal grouped data are considered, and parameter estimates are provided for the three methods described in the previous paragraph. In Section 2.2, the proposed methods are applied to bivariate grouped data and extended to multivariate normal grouped data. Standard errors for the EM and MCEM mean estimates are presented in Section 2.3. Section 3 deals with numerical applications. In Section 3.1,
the proposed methods are applied to the well-known Galton data \cite{NaturalInheritance}.
Simulation studies for univariate and bivariate normal grouped data are described in Section 3.2. Finally, in Section 4, results and conclusions are discussed.    

\section{Methods}

\subsection{Univariate normal grouped data }
For simplicity, this section starts with estimating the parameters for the univariate normal grouped data.
\subsubsection{Exact MLE}
It is assumed that the unobserved data $x_1,x_2,\dots,x_n$ come from a normal distribution with parameters $\theta = ( \mu, \sigma)$ and denoted by $N(\mu,\sigma)$.
Let $f(x;\theta)$ be the density function of $N(\mu,\sigma)$. According to $k+1$ pre-established partitioned points $a_0<a_1< \dots<a_{k-1}<a_{k}$, let $n_i$ be the number of observations that fall into the interval $\mathcal{X}_i = [a_{i-1},a_i)$ for $1\leq i \leq k$, $a_0=-\infty$ and $a_{k}=+\infty$.
Furthermore, it is assumed that the observed data $y=\{n_1,\dots,n_{k}\}$ follow a multinomial distribution with  $n=\sum_{i=1}^{k}n_i$ draws over $k$ categories (intervals), with the probability of being in category $i$ equal to $\frac{P_i(\theta)}{P(\theta)}$, where
$$P_i(\theta)=\int \limits_{a_{i-1}}^{a_i} f(x;\theta)dx,$$

\noindent with $P(\theta)=\sum_{i=1}^{K} P_i(\theta) = 1$. Therefore, the log-likelihood function for the observed data $y$ (also called the incomplete data log-likelihood) can be written as:
\begin{equation} 
\log L(\theta)=\sum_{i=1}^{k} n_i \log P_i(\theta) + C. 
\label{EQ3}
\end{equation}

Let $\phi(\cdot)$ and $\Phi(\cdot)$ be the density and the cumulative distribution function (CDF); respectively; of a standard normal distribution. Therefore, the density of $N(\mu,\sigma)$ can be written as:
$$f(x;\mu,\sigma)=\frac{1}{\sigma}\phi(\frac{x-\mu}{\sigma}),$$
where $-\infty<\mu<\infty$ and $\sigma>0$. 
By applying the reparametrization 
$\theta_1=\frac{\mu}{\sigma}$
and
$\theta_2=\frac{1}{\sigma}$, 
the parameters are changed from $\theta=(\mu,\sigma)$ to $\theta=(\theta_1,\theta_2)$. Now let the CDF of $N(\mu,\sigma)$ be  $\Phi(\theta_2t-\theta_1)$. Then the log-likelihood in (\ref{EQ3}) can be written as a function of $\theta_1$ and $\theta_2$ as follows (see also \cite{Xia2009}):
\begin{eqnarray}
\log L(\theta) &=&n_1\ln\Big[\Phi(\theta_2 a_1-\theta_1)\Big]+n_{k}\ln\Big[1-\Phi(\theta_2a_{k-1}-\theta_1) \Big]+ \nonumber \\
& & \sum_{i=2}^{k-1} n_i\ln \Big[\Phi(\theta_2a_i-\theta_1)-\Phi(\theta_2a_{i-1}-\theta_1) \Big] + C. 
\label{Eq:loglik reparam}
\end{eqnarray}

The parameter estimates $\hat{\theta}_1$ and $\hat{\theta}_2$ can be obtained by maximizing (\ref{Eq:loglik reparam}) with respect to $\theta = \{\theta_1,\theta_2\}$ using  Newton-Raphson numerical methods such as those implemented in the \textit{optim()} function in R. 

\subsubsection{Parameter estimation via the EM algorithm}
In a similar manner to \cite{Mclachlan2007,McLachlan1988,Chanseok2006}, to find $\hat{\theta}$ that maximizes $\log L(\theta)$ in (\ref{EQ3}) within the EM framework, the vector of $x_i=(x_{i1},x_{i2},\dots,x_{in_i})^T$, for $i=1,\dots,k$, should be introduced as missing (unobservable) data. In fact,
for each interval $\mathcal{X}_i=[a_{i-1},a_i)$, $x_i$ consists of $n_i$ independent unobservable data points falling into that interval. Hence, the complete-data vector can be written as $w=(y^T, x_1^T,\dots,x_{k}^T)^T$. Furthermore, given $y$, each $x_{il}$ has a density function  $f(x_{il}|y) \equiv\frac{f(x_{il};\theta)}{P_i(\theta)}$ for $l=1,\ldots,n_i$ and $i=1,\dots,k$.
Therefore, the complete-data likelihood of can be written as
\begin{eqnarray*}
L_c(\theta) &\equiv& f(w;\theta) \\
  &=& f(x|y;\theta) p(y;\theta) \\
 &=& \prod_{i=1}^{k}   \prod_{l=1}^{n_i} \frac{f(x_{il};\theta)} {Pi(\theta)} \times \prod_{i=1}^{k} (P_i(\theta))^{n_i} \times C \\
 & \propto & \prod_{i} \prod_{l}f(x_{il};\theta),
\end{eqnarray*}
and its corresponding log-likelihood as
\begin{equation}\label{EQ4}
\log L_c(\theta)\equiv \sum_{i=1}^{k}\sum_{l=1}^{n_i} \log f(x_{il};\theta)+C \mbox{.}
\end{equation}
Using (\ref{EQ4}), the EM algorithm can be used to iteratively estimate the parameters in $\theta$. The following describes the E and M steps of the proposed EM approach.

\vspace{0.5cm}
\noindent \textbf{E-Step}:\\
The E-step calculates the expectation of the complete-data log-likelihood in (\ref{EQ4}) conditional on $y$ and the current parameter estimates ($\theta^{(p)}$). Disregarding the constant term, the expectation of the $\log L_c(\theta)$ conditional on $y$ and $\theta^{(p)}$ is given by:
\begin{eqnarray*}
Q(\theta,\theta^{(p)})\equiv E_{\theta^{(p)}} \Big [ \log L_c(\theta)|y \Big ] = \sum_{i=1}^k n_i E_{\theta^{(p)}} \Big [ \log f(X;\theta)|X \in \mathcal{X}_i \Big ], 
\end{eqnarray*}
where the expectation is taken with respect to the density $\frac{f(x;\theta^{(p)})}{P_i(\theta^{(p)})}$.

Therefore, for the normally distributed grouped data, we can write:
\begin{eqnarray*}
Q(\theta,\theta^{(p)})= -\frac{1}{2}n \{\log(2\pi)+\log \sigma^2 \}-\frac{1}{2}\sigma^2 \sum_{i=1}^{k} n_i E_{\theta^{(p)}} \Bigg \{(X-\mu)^2|X \in \mathcal{X}_i \Bigg \}\mbox{.}
\end{eqnarray*}

\noindent \textbf{M-Step}:

The M-step of the EM algorithm maximizes $Q(\theta,\theta^{(p)})$ with respect to $\theta$ at iteration $p+1$ to produce new estimates $\theta^{(p+1)}=(\mu^{(p+1)},\sigma^{(p+1)})^T$.  
By using the idea of interchanging the differentiation and the expectation (the Leibniz integral rule), $Q(\theta,\theta^{(p)})$ can be differentiated with respect to $\theta=(\mu,\sigma)$ to obtain the following updated estimates:
\begin{equation} \label{EQ5}
    \mu^{(p+1)}=\frac{\sum_{i=1}^{k} n_i E_{\theta^{(p)}}(X|X \in \mathcal{X}_i)}{n} 
\end{equation}
\noindent and
\begin{equation} \label{EQ6}
    \sigma^{2(p+1)}=\frac{\sum_{i=1}^{k} n_i E_{\theta^{(p)}}\Big[(X-\mu^{(p+1)})^2|X \in \mathcal{X}_i\Big]}{n},
\end{equation}
where $n=\sum_{i=1}^{k}n_i$. The derivation of the expectations in (\ref{EQ5}) and (\ref{EQ6}) can be found in Section \ref{Sec:Univariate} of the Appendix.
\subsubsection{Parameter estimation via the MCEM Algorithm}
Calculating the exact form of the expectations in (\ref{EQ5}) and (\ref{EQ6}) can be seen as a tedious and difficult task. In this case, an alternative way is to use the Monte-Carlo EM (MCEM) algorithm, in which the required expectations are replaced with an average over simulations \cite{Mclachlan2007, Chanseok2006}. 
The unobserved data $x=(x_1,\dots,x_k)$ can be simulated from the distribution  $\frac{f(x;\theta^{(p)})}{P_i(\theta^{(p)})}$ over each specific interval. Now, considering $M$ as the number of observations generated for each interval in the Monte-Carlo simulation, the simulated sample for the $i$-th interval can be written  as $(x_{i1},\dots,x_{iM})$, and the MCEM updates are:
    
\begin{equation} \label{EQ7}
    \mu^{(p+1)}=\frac{1}{n} \sum_{i=1}^{k}n_i \frac{1}{M} \sum_{m=1}^{M}x_{im} \nonumber
\end{equation}
and 
\begin{equation} \label{EQ8}
     \sigma^{2(p+1)}=\frac{1}{n} \sum_{i=1}^{k}n_i \frac{1}{M} \sum_{m=1}^{M}(x_{im}-\mu^{(p+1)})^2 \text{,}\nonumber
\end{equation}
where $n=\sum_{i=1}^{k}n_i$.

\subsection{Bivariate and Multivariate Normal grouped data}
\subsubsection{Exact MLE for Bivariate Normal}
\begin{center}
\begin{table}
\begin{tabular}{|c|c|c|c|c|c|}\hline
$x_1\backslash x_2$  & $[b_0,b_1)$& $[b_1,b_2)$&$\cdots$& $[b_{s-1},b_s)$&Total \\ \hline
 $[a_0,a_1)$ &$n_{11}$&$n_{12}$& $\cdots$ &$n_{1s}$&$n_{1.}$\\\hline
 $[a_1,a_2)$ &$n_{21}$&$n_{22}$&$\cdots$ & $n_{2s}$&$n_{2.}$\\\hline
 \vdots &\vdots&\vdots&\vdots&\vdots&\vdots\\\hline
$[a_{r-1},a_r)$ &$n_{r1}$&$n_{r2}$&$\cdots$&$n_{rs}$&$n_{r.}$\\\hline
 Total &$n_{.1}$&$n_{.2}$&$\cdots$&$n_{.s}$&$n$ \\ \hline
 \end{tabular}
 \caption{Bivariate grouped data representation.}\centering
\label{table:Bivariate Grouped data}
\end{table}
\end{center}
The derivation of the exact MLE for bivariate normal grouped data is much like that for the univariate case, except that the multinomial probabilities depend on the bivariate normal CDF calculated over rectangles instead of intervals. The probability of a bivariate random variable $X=(X_1,X_2)$ belonging to a rectangale $\mathcal{X}_1\times \mathcal{X}_2$ of the form $[a_{i-1},a_i) \times [b_{j-1}, b_j)$; for $i=1,\dots,r$ and $j=1,\dots,s$, is 
\begin{eqnarray*}\label{EQ13}
P_{ij}(\theta)&\equiv&P(a_{i-1} \leq X_1 < a_i,b_{j-1} \leq X_2 < b_j )=\\
&=&\int \limits_{a_{i-1}}^{a_i} \int \limits_{b_{j-1}}^{b_j} f(x_1,x_2;\theta)dx_1dx_2\\
&=&F_{\theta}(a_i,b_j)-F_{\theta}(a_{i-1},b_j)-F_{\theta}(a_i,b_{j-1})+F_{\theta}(a_{i-1},b_{j-1}),
\end{eqnarray*} 
where $f(.;\theta)$ and $F_{\theta}(.)$ are the bivariate normal density function and cumulative distribution function, respectively; with parameters $\theta=(\mu_{x_1},\mu_{x_2},\sigma_{x_1},\sigma_{x_2},\rho)$.

For each rectangle (or cell in Table \ref{table:Bivariate Grouped data}), the frequencies $n_{ij}$, for $i=1,\dots,r$ and $j=1,\dots,s$), are known, and therefore the following multinomial likelihood can be assumed for them:
$$L(\theta)=\frac{n!}{\prod_{i=1}^{r} \prod_{j=1}^{s} n_{ij}}\prod_{i=1}^{r} \prod_{j=1}^{s} \Big[\frac{P_{ij}(\theta)}{P(\theta)}\Big]^{n_{ij}} \mbox{,}$$
where $n=\sum_{i=1}^{r} \sum_{j=1}^{s} n_{ij}$ and $P(\theta)=\sum_{i=1}^{r} \sum_{j=1}^{s} P_{ij}(\theta)=1$. Hence, the exact log-likelihood function is:
\begin{align}
&\log L(\theta)=\sum_{i=1}^{r} \sum_{j=1}^{s} n_{ij} \log P_{ij}(\theta)+C  \nonumber \\
&=\sum_{i=1}^{r} \sum_{j=1}^{s} n_{ij}\log \Big[F_{\theta}(a_{i},b_{j})-F_{\theta}(a_{i-1},b_{j})-F_{\theta}(a_{i},b_{j-1})+F_{\theta}(a_{i-1},b_{j-1}) \Big]+C \label{EQ15}
\end{align}

To find the MLEs of the parameters in $\theta$, the log-likelihood function in (\ref{EQ15}) is maximized using numerical methods implemented by the \textit{nlm()} function in R.
\subsubsection{Parameter Estimation via the EM Algorithm}
Extending the ideas of the univariate case, the goal is to maximize the exact log-likelihood for bivariate grouped data (see Equation  (\ref{EQ15})); using the EM approach. 
Therefore, the first step is to introduce $x$ as missing observations in array form as:
$$x =\big \{(x_{1ik},x_{2jk}) \;\mbox{for}\; i=1,\ldots,r; \;  j=1,\ldots, s; \; k=1,\ldots,n_{ij} \big \}.$$
Then the complete-data $w=\big\{ y,x\big \}$ can be defined over the rectangles, and their log-likelihood can be written as:
\begin{align}
\log L_c(\theta)&=\log L(\theta)+\sum_{i=1}^{r} \sum_{j=1}^{s} \sum_{k=1}^{n_{ij}} \log \frac{f(x_{1ik},x_{2jk};\theta)}{P_{ij}(\theta)} \nonumber\\
&= \sum_{i=1}^{r}\sum_{j=1}^{s}  n_{ij} \log P_{ij}(\theta)+C+\sum_{i=1}^{r}\sum_{j=1}^{s}  \sum_{k=1}^{n_{ij}} \log \frac{f(x_{1ik},x_{2jk};\theta)}{P_{ij}(\theta)} \nonumber\\
&=\sum_{i=1}^{r}\sum_{j=1}^{s} \sum_{k=1}^{n_{ij}} \log f(x_{1ik},x_{2jk};\theta)+C 
\label{EQ15-1}
\end{align}
The following presents the proposed E and M steps of the EM algorithm.

\vspace{0.5cm}
\noindent \textbf{E-Step}:

The E-step calculates the expected value of (\ref{EQ15-1}) given $y$ and the current $\theta^{(p)}$, that is,
\begin{equation}\label{EQ15-2}
 Q(\theta,\theta^{(p)})\equiv\sum_{i=1}^{r} \sum_{j=1}^{s}n_{ij}Q_{ij}(\theta,\theta^{(p)}) \mbox{,} 
\end{equation}
where
$$Q_{ij}(\theta,\theta^{(p)})=E_{\theta^{(p)}}\Bigg\{\log f\Big((X_1,X_2);\theta\Big)|(X_1,X_2)\in (\mathcal{X}_{i1} \times \mathcal{X}_{2j}) 
\Bigg \} \text{,}$$
with the expectation taken with respect to the density $\frac{f((x_1,x_2);\theta^{(p)})}{P_{ij}(\theta^{(p)})}$. 

\noindent \textbf{M-Step}:

The M-step aims to find the parameter updates that maximize (\ref{EQ15-2}). Using a similar framework as in Section 2.1.2, the results are:
\begin{equation}\label{EQ16}
\mu_{x_1}^{(p+1)}=\frac{\sum_{i=1}^{r}\sum_{j=1}^{s}n_{ij}E_{\theta^{(p)}}\bigg(X_{1i}|(X_1,X_2)\in (\mathcal{X}_{i1} \times \mathcal{X}_{2j})\bigg)}{n} 
\end{equation}

\begin{equation}\label{EQ17}
\mu_{x_2}^{(p+1)}=\frac{\sum_{i=1}^{r}\sum_{j=1}^{s}n_{ij}E_{\theta^{(p)}}\bigg(X_{2j}|(X_1,X_2)\in (\mathcal{X}_{i1} \times \mathcal{X}_{2j})\bigg)}{n} 
\end{equation}

\begin{equation}\label{EQ18}
\sigma_{x_1}^{2(p+1)}=\frac{\sum_{i=1}^{r}\sum_{j=1}^{s}n_{ij}E_{\theta^{(p)}}\bigg((X_{1i}-\mu^{(p+1)}_{x_1})^2|(X_1,X_2)\in (\mathcal{X}_{i1} \times \mathcal{X}_{2j})\bigg)}{n} 
\end{equation}

\begin{equation}\label{EQ19}
\sigma_{x_2}^{2(p+1)}=\frac{\sum_{i=1}^{r}\sum_{j=1}^{s}n_{ij}E_{\theta^{(p)}}\bigg((X_{2j}-\mu^{(p+1)}_{x_2})^2|(X_1,X_2)\in (\mathcal{X}_{i1} \times \mathcal{X}_{2j})\bigg)}{n} 
\end{equation}

\begin{equation}\label{EQ20}
\rho^{(p+1)}=\frac{\sum_{i=1}^{r}\sum_{j=1}^{s} n_{ij}E_{\theta^{(p)}}\bigg((X_{1i}-\mu^{(p+1)}_{x_1})(X_{2j}-\mu^{(p+1)}_{x_2})|(X_1,X_2)\in (\mathcal{X}_{i1} \times \mathcal{X}_{2j})\bigg)}{n} 
\end{equation}
The expectations in (\ref{EQ16}) to (\ref{EQ20}) are the moments of a truncated bivariate normal distribution  ($\frac{f(x_1,x_2;\theta^{(p)})}{P_{ij}(\theta^{(p)})}$) and, therefore, the results of \cite{Manjunath2021} are used to calculate each one of them (for details, see Section \ref{Sect::Multivariate} of the Appendix).

\subsubsection{MCEM for bivariate grouped data}
The MCEM algorithm can be used to replace the expectations in (\ref{EQ16}) to (\ref{EQ20}) by the average of simulated values. 
That means that $M$ random samples of $(X_1,X_2)$ are simulated from the density $\frac{f((x_1,x_2);\theta^{(p)})}{P_{ij}(\theta^{(p)})}$ over the rectangles, and then their averages are used to replace  the expectations in the EM parameter updates, obtaining the following MCEM-based parameter estimates:
\begin{eqnarray*}\
&&\mu_{x_1}^{(p+1)}=\frac{\sum_i \sum_j n_{ij} \frac{1}{M} \sum_{h=1}^{M} x_{1ih}}{n}  \mbox{,}\\ 
&&\mu_{x_2}^{(p+1)}=\frac{\sum_i \sum_j n_{ij} \frac{1}{M} \sum_{h=1}^{M} x_{2jh}}{n}  \mbox{,}\\ 
&&\sigma_{x_1}^{2(p+1)}=\frac{\sum_i \sum_j n_{ij} \frac{1}{M} \sum_{h=1}^{M} (x_{1ih}-\mu^{(p+1)}_{x_1})^2}{n} \mbox{,} \\ 
&&\sigma_{x_2}^{2(p+1)}=\frac{\sum_i \sum_j n_{ij} \frac{1}{M} \sum_{h=1}^{M} (x_{2jh}-\mu^{(p+1)}_{x_2})^2}{n} \mbox{, and}  \\ 
&&\rho^{(p+1)}=\frac{\sum_i \sum_j n_{ij} \frac{1}{M} \sum_{h=1}^{M} (x_{1ih}-\mu^{(p+1)}_{x_1})(x_{2jh}-\mu^{(p+1)}_{x_2})}{n} \mbox{.} 
\end{eqnarray*}
\subsubsection{Extension of EM and MCEM to Multivariate Normal Grouped Data}
By extending the ideas of univariate and bivariate normal grouped data, it is possible to find the parameter estimates (mean vector and covariance matrix) for multivariate normal grouped data using a matrix notation.
Let $(x_1,\dots,x_d)$ be an unobservable vector arising from a $d$-dimensional multivariate normal distribution with parameters $\Theta=(\underline{\mu},\Sigma)$. 
Consider $r_1 r_2 \cdots r_d$ as the number of $d$-dimensional surfaces of the form  $\mathcal{X}_{1i_1}\times \mathcal{X}_{2i_2} \times \dots \times \mathcal{X}_{di_d}$ = $[a_{1i_1-1},a_{1i_1}]\times[a_{2{i_2-1}},a_{2i_2}] \times \dots \times [a_{di_{d}-1},a_{di_d}]$ for $i_1=1,\dots,r_1$; $i_2=1,\dots,r_2$; $ \dots,i_d=1,\dots,r_d$. Let $n_{i_1,\dots,i_d}$ be the observed number (count) of data points falling in each surface. These observed counts form a multinomial likelihood as follows:
$$L(\Theta)\equiv \frac{n!}{\prod_{i_1=1}^{r_1} \dots \prod_{i_d=1}^{r_d}(n_{i_1,i_2,\dots,i_d})!}\prod_{i_1=1}^{r_1}\dots \prod_{i_d=1}^{r_d}\Big(\frac{P_{i_1,i_2,\dots,i_d}(\Theta)}{P(\Theta)}\Big)^{n_{i_1,\dots,i_d}} $$
where $n=\sum_{i_1}\dots\sum_{i_d}n_{i_1,\dots,i_d}$,

$$P_{i_1,i_2,\dots,i_d}(\Theta) \equiv \int\limits_{a_{1i_1-1}}^{a_{1i_1}}\dots \int\limits_{a_{di_d-1}}^{a_{di_d}}f(x_1,\dots,x_d)dx_d\dots dx_1,$$ 
and
$$P(\Theta)\equiv \sum_{i_1=1}^{r_1}\dots\sum_{i_d=1}^{r_d}P_{i_1,i_2,\dots,i_d}(\Theta)=1 \mbox{,}$$
with $f(x_1,\dots,x_d)$ being the probability density function of a multivariate normal distribution.
Representing the observed data as $y=\big \{n_{i_1,\dots,i_d} \; \mbox{for} \;  i_1=1,\dots,r_1; \dots; i_d=1,\dots,r_d \big \}$, the goal is to maximize 
$$\log L(\Theta)=\sum_{i_1}\dots\sum_{i_d} n_{i_1,\dots,i_d} P_{i_1,\dots,i_d}(\Theta)+C,$$

\noindent with respect to $\Theta$ using the EM framework.
Let $x = \big \{ (x_{1i_1k},x_{2i_2k},\dots,x_{di_dk}) \; \mbox{for} \;  i_1=1,\dots,r_1; \dots; i_d=1,\dots,r_d;k=1,2,\dots,n_{i_1,\dots,i_d} \big \}$ be the missing vectors of observations. Thus, considering the complete data as $w=\big \{y,x \big \}$, the complete-data log-likelihood function can be written as:  
\begin{eqnarray*}
\log L_c(\Theta)&=&\log L(\Theta)+\sum_{i_1=1}^{r_1}\dots \sum_{i_d=1}^{r_d} \sum_{k=1}^{n_{i_1,\dots,i_d}} \log \frac{f\Big((x_{1i_1k},x_{2i_2k},\dots,x_{di_dk});\Theta\Big)}{P_{i_1,\dots,i_d}(\Theta)}\\
&=& \sum_{i_1=1}^{r_1}\dots\sum_{i_d=1}^{r_d}  n_{i_1,\dots,i_d} \log P_{i_1,\dots,i_d}(\Theta)+C\\
&+&\sum_{i_1=1}^{r_1}\dots\sum_{i_d=1}^{r_d}  \sum_{k=1}^{n_{i_1,\dots,i_d}} \log \frac{f\Big((x_{1i_1k},x_{2i_2k},\dots,x_{di_dk});\Theta\Big)}{P_{i_1,\dots,i_d}(\Theta)}\\
&=&\sum_{i_1=1}^{r_1}\dots\sum_{i_d=1}^{r_d} \sum_{k=1}^{n_{i_1,\dots,i_d}} \log f\Big((x_{1i_1k},x_{2i_2k},\dots,x_{di_dk});\Theta\Big)+C
\end{eqnarray*}
The E-step and M-step of the EM algorithm are described as follows.

\vspace{0.5cm}
\noindent \textbf{E-Step}:

The E-step calculates:
\begin{eqnarray*}
Q(\Theta,\Theta^{(p)})=\sum_{i_1=1}^{r_1}\dots \sum_{i_d=1}^{r_d}n_{i_1,\dots,i_d}Q_{i_1,\dots,i_d}(\Theta,\Theta^{(p)}),
\end{eqnarray*}

\noindent where
\begin{eqnarray*}
Q_{i_1,\dots,i_d}(\Theta,\Theta^{(p)})=E_{\Theta^{(p)}}\Bigg\{\log f\Big((X_1,\dots,X_d);\Theta \Big)\Big |
(X_1,\dots,X_d)\in (\mathcal{X}_{1i_1} \times \dots \times \mathcal{X}_{di_d}) 
\Bigg \} \text{.}
\end{eqnarray*}

Hence,
\begin{eqnarray*}
&&Q(\Theta,\Theta^{(p)})=\sum_{i_1}\dots \sum_{i_d}n_{i_1,\dots,i_d}E_{\theta^{(p)}}\bigg\{\Big[ -\frac{d}{2}\log (2\pi)+\frac{1}{2}\log(|\Sigma|^{-1})\\
&&-\frac{1}{2}(x_i-\mu)^T\Sigma^{-1}(x_i-\mu) \Big] \Big|(X_1,\dots,X_d)\in (\mathcal{X}_{1i_1} \times \dots \times \mathcal{X}_{di_d}) \bigg\}\\
&&=\sum_{i_1}\dots \sum_{i_d}n_{i_1,\dots,i_d}\Bigg\{ -\frac{d}{2}\log (2\pi)
+\frac{1}{2}\log(|\Sigma|^{-1})\\
& &- \frac{1}{2}Tr\Bigg[\Sigma^{-1} E_{\Theta^{(p)}}\Big((x_i-\mu)(x_i-\mu)^T  \Big|(X_1,\dots,X_d)\in (\mathcal{X}_{1i_1} \times \dots \times \mathcal{X}_{di_d})\Big) \Bigg] \Bigg\}\mbox{.}
\end{eqnarray*} 

\noindent \textbf{M-Step}:

The M-step maximizes $Q(\Theta,\Theta^{(p)})$ w.r.t $\Theta$, obtaining:

\begin{equation}\label{EQ21}
\underline{\mu}^{(p+1)}=\frac{1}{n}\sum_{i_1=1}^{r_1}\dots\sum_{i_d=1}^{r_d}n_{i_1,\dots,i_d} 
E_{\Theta^{(p)}}\Bigg \{X_i \Big|(X_1,\dots,X_d)\in (\mathcal{X}_{1i_1} \times \dots \times \mathcal{X}_{di_d})\Bigg\}
\end{equation}

\begin{align}
&\Sigma^{(p+1)}=\frac{1}{n}\sum_{i_1=1}^{r_1}\dots\sum_{i_d=1}^{r_d}n_{i_1,\dots,i_d} \times \nonumber \\
&E_{\Theta^{(p)}}\Bigg \{\Big((X_i-\mu^{(p+1)})(X_i-\mu^{(p+1)})^T \Big) \Big|(X_1,\dots,X_d)\in (\mathcal{X}_{1i_1} \times \dots \times \mathcal{X}_{di_d})\Bigg\}. \label{EQ22}
\end{align}

The expectations in (\ref{EQ21}) and (\ref{EQ22}) are the moments of a truncated multivariate normal $\frac{f(x_1,\dots,x_d;\Theta)}{P_{i_1,\dots,i_d}(\Theta)}$ and as in the bivariate case, the results in \cite{Manjunath2021} are used to calculate these moments, as shown in Appendix \ref{Sect::Multivariate}.

The calculations of these expectations for the multivariate normal case, particularly for $d>2$, are complex and error-prone. To avoid such calculations, an alternative approach is to use the MCEM algorithm. The MCEM approach first simulates $M$ multivariate random samples of $X=(X_1,\dots,X_d)$ from the density  $\frac{f(x_1,\dots,x_d;\Theta)}{P_{i_1,\dots,i_d}(\Theta)}$ over all surfaces and then replaces the expectations in (\ref{EQ21}) and (\ref{EQ22})  with the averages of the simulated sample vectors obtaining the following parameter updates:
\begin{equation*}
\underline{\mu}^{(p+1)}=\frac{1}{n}\Bigg[\sum_{i_1}\dots \sum_{i_d} n_{i_1,\dots,i_d} \frac{1}{M}\sum_{h=1}^{M}x_{hi}\Bigg]
\end{equation*}
\noindent and
\begin{equation*}
\Sigma^{(p+1)}=\frac{1}{n}\Bigg[\sum_{i_1}\dots \sum_{i_d} n_{i_1,\dots,i_d} \frac{1}{M}\sum_{h=1}^{M}(x_{hi}- \mu^{(p+1)})(x_{hi}-\mu^{(p+1)})^T \Bigg].
\end{equation*}
\subsection{Standard Errors for the  EM and MCEM mean Estimates}
Following the ideas in Chapter 4 of \cite{Mclachlan2007}, standard errors for the EM estimates for grouped data can be obtained using an approximation of the observed information matrix, which is called the empirical observed information matrix, $I_{e,g}$. For the univariate grouped data, $I_{e,g}$ can be calculated as: 
\begin{equation}
\label{IE_Univariate}
    I_{e,g}(\hat{\theta};y)=\sum_{i=1}^{r}n_i s_i(\hat{\theta})s^T_i(\hat{\theta})-n \Bar{s}(\hat{\theta})\Bar{s}^T(\hat{\theta}),
\end{equation}
where $\Bar{s}(\hat{\theta})=\frac{1}{n}\sum_{i=1}^{r}n_is_i(\hat{\theta})$, $s_i(\hat{\theta})=\frac{\partial Q_i(\theta,\hat{\theta})}{\partial \theta}|_{\theta=\hat{\theta}}$, and $\hat{\theta}$ contains the EM estimates. 

Similarly, the empirical observed information matrix for multivariate grouped data is as follows: 
\begin{equation}
\label{IE_Multivariate}
    I_{e,g}(\hat{\Theta};y)=\sum_{i_1=1}^{r_1}\dots\sum_{i_d=1}^{r_d}    n_{i_1,\dots,i_d} s_{i_1,\dots,i_d}(\hat{\Theta})s^T_{i_1,\dots,i_d}(\hat{\Theta})-n \Bar{s}(\hat{\Theta})\Bar{s}^T(\hat{\Theta})
\end{equation}
where $\displaystyle s_{i_1,\dots,i_d}(\hat{\Theta})=\frac{\partial Q_{i_1,\dots,i_d}(\Theta,\hat{\Theta})}{\partial \Theta}|_{\Theta=\hat{\Theta}}$.

The inverse of $I_{e,g}$ demonstrates an approximation of the covariance matrix of the EM estimates, with the diagonal containing the standard errors.  

For our study, we calculate the standard error for the EM estimates of $\mu$ and $\underline{\mu}$ using equations (\ref{IE_Univariate}) and (\ref{IE_Multivariate}), respectively, and fixing the variance-covariance parameter values to ones obtained by the EM algorithm. 
Using the notation from the previous sections, we can show that the score function for $\mu$ for univariate grouped data is: 
\begin{equation*}
    s_i(\hat{\mu},\hat{\sigma}^2) = \frac{1}{\hat{\sigma}^2} E\Big\{(X-\hat{\mu})|X\in \mathcal{X}_i \Big\}, 
\end{equation*}
and for the multivariate case is:
\begin{equation*}
    s_{i_1,\dots,i_d}(\hat{\underline{\mu}},\hat{\Sigma}^2)= E\Big \{(X-\hat{\underline{\mu}})|X\in (\mathcal{X}_{1i_1} \times \dots \times \mathcal{X}_{di_d}) \Big\}^T\hat{\Sigma}^{-1} \mbox{.}
\end{equation*}
Using the previous score functions, we also obtain standard errors for the mean MCEM estimates using the Louis' approach \cite{louis1982finding} as described in Chapter 6 of \cite{Mclachlan2007} with all expectations replaced by the averages of observations simulated using the final MCEM estimates. 


We use the standard errors ($se$) proposed above to construct 95\% confidence intervals of the form: $\hat{\mu}\pm 1.96 se(\hat{\mu})$. 

\section{Results}
\subsection{Galton Data}
The Galton dataset was first introduced by Francis Galton in 1886 \cite{HANLEY2004,NaturalInheritance} and consists of a two-way frequency table containing the number of parents and children falling into different possible height intervals.
The individual height observations are not available; only the frequencies (grouped data) are available. Moreover, for each interval, the midpoints (as the averages of the lower and upper limits of the intervals) are also available. 
This data set is a well-known example of normally distributed grouped data. The Galton data are electronically and publicly available in the R package \textit{HistData}. 
In this study, each of the variables (parent's height and child's height) was first analyzed separately as univariate normal grouped  data before considering the bivariate case. 
The results are provided in Sections 3.1.1 and 3.1.2 respectively. 
\subsubsection{Univariate Case}\label{secuni}

First, the exact MLE of the parameters with the assumption of normal distribution of both parent height and child height data were obtained using the approach described in Section 2.1.1. As described, for the grouped data, the exact likelihood estimation was conducted numerically using the R functions \textit{optim()} and \textit{nlm} (L-BFGS-B method); the results are shown in Table \ref{Galton Univariate} under Exact MLE. Note that the numerical maximization of the exact likelihood is highly sensitive to initial values.
The parameter estimates using the EM algorithm to maximize the exact likelihood were then found, along with those  using the MCEM algorithm. The results for both EM and MCEM algorithms are also presented in Table \ref{Galton Univariate}.
As can be expected by the convergence properties of the EM algorithm \cite{Mclachlan2007}, its estimates were close to those obtained by direct maximization of the exact likelihood (mean absolute relative difference (MARD) across parameters = $0.005672\%$). 
The MCEM estimates were also close to the Exact MLE results (MARD = $0.020222\%$), but not as close as the EM results, which was also expected from the properties of the MCEM \cite{Mclachlan2007}. 

\begin{table}[ht]
\centering
\small
\caption{Estimates of the mean (with its standard errors (se) for EM and MCEM) and variance (Var) of parent and child height variables (considering the univariate case) from the Galton data using the three proposed methods.}
\begin{tabular}{|rrrrr|}
  \hline
 Method& Mean parent (se) & Mean child (se)& Var parent & Var child \\ 
  \hline
 Exact MLE & 68.30030 & 68.09834  & 3.24432 & 6.50945 \\ 
  EM & 68.30026 (0.03818) & 68.09834 (0.05232) & 3.24482 & 6.50971 \\ 
  MCEM & 68.30070 (0.05992) & 68.09600 (0.08435) & 3.24312 & 6.50763 \\ 
   \hline
\end{tabular}
\label{Galton Univariate}
\end{table}

\subsubsection{Bivariate Case}\label{secBiv}
In this case, the Galton data were considered as bivariate grouped data and the methods proposed in Section 2.2 were used to find the parameter estimates. The results for all five parameters (including mean of parents, mean of children, variance of parents, variance of children, and correlation of heights between parents and children) are shown in Table \ref{BivariateGalton}. Note that as mentioned in Section 2.2.1, for parameter estimates using the exact MLE method for bivariate data, the \textit{nlm()} and \textit{optim()} functions in R were used. The EM estimates were closest to those from the exact MLE method, with mean absolute relative difference over the five parameters of $0.0012\%$.
 \begin{table}[ht]
\centering
\small
\caption{Estimates of mean, variance (Var) and correlation (Corr) parameters for bivariate Galton data using the three proposed methods. Standard errors (se) for the mean EM and MCEM estimates are also provided.} 
\begin{tabular}{|rrrrrr|}
  \hline
Method & Mean parent & Mean child & Var parent & Var child & Corr \\ 
  \hline
 Exact MLE & 68.300475 & 68.098651 & 3.243895 & 6.513746 & 0.470162 \\ 
  EM & 68.300495(0.059656) & 68.098736(0.084259) & 3.243960 & 6.513621 & 0.470171 \\ 
  MCEM & 68.302157(0.058073) & 68.098961(0.070917) & 3.248326 & 6.514850 & 0.469763 \\ 
   \hline
\end{tabular}

\label{BivariateGalton}
\end{table}

\subsection{Simulation Studies}
In this section, the parameter estimation methods for normally distributed grouped data are applied to simulated data for both the univariate and bivariate cases. Resulting tables and plots are displayed at the end of this manuscript. 

\subsubsection{Univariate Simulation}
In this study, we conducted simulations on 15 different scenarios obtained by varying the sample size $n$ (50, 100, 300, 600 and 1000) and the number of equal-sized intervals (or bins, $k=8, 15,$ and 30). For each scenario, 500 univariate datasets (in total 7500 datasets) are simulated. All simulated data are from a univariate normal distribution with parameters $\mu=68$ and $\sigma=2.5$ ($\sigma^2=6.25)$. Moreover, according to \cite{Booth1999} and \cite{McCulloch1997}, as the number of MCEM iterations for the univatiate data was between 10 to 30, we fix the number of Monte-Carlo simulations for MCEM estimates to $m=1000$. 

\begin{figure}[ht]
\centering
\includegraphics[width=1.15\textwidth]{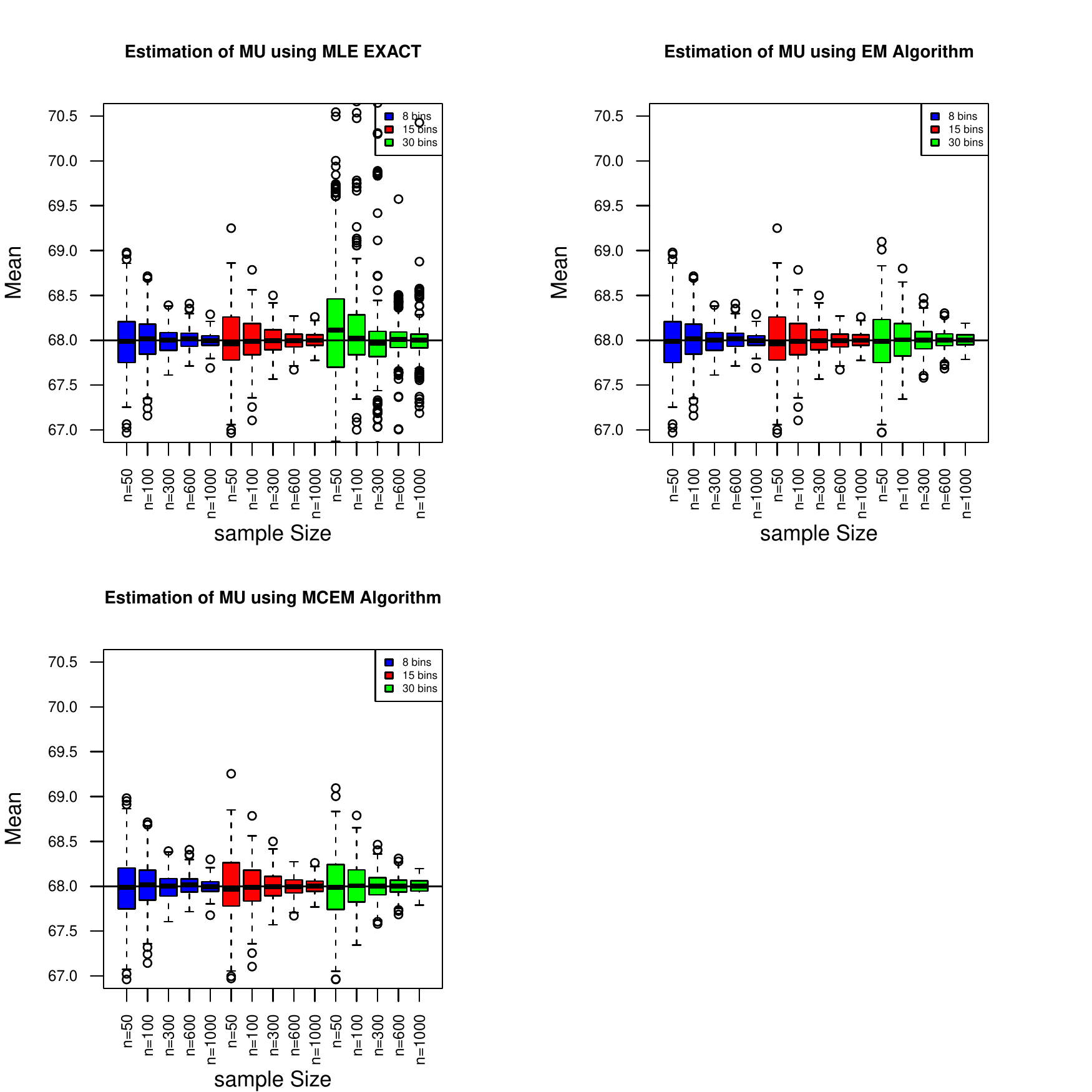}
\caption{\textit{Simulation results: univariate case.} Mean estimates for $k= 8
$, 15 and 30 intervals (bins) for sample sizes $n=50,100,300,600,1000$. True mean  value $\mu=68$.}\label{fig:MU}
\end{figure}

\begin{figure}[ht]
\centering
\includegraphics[width=1.15\textwidth]{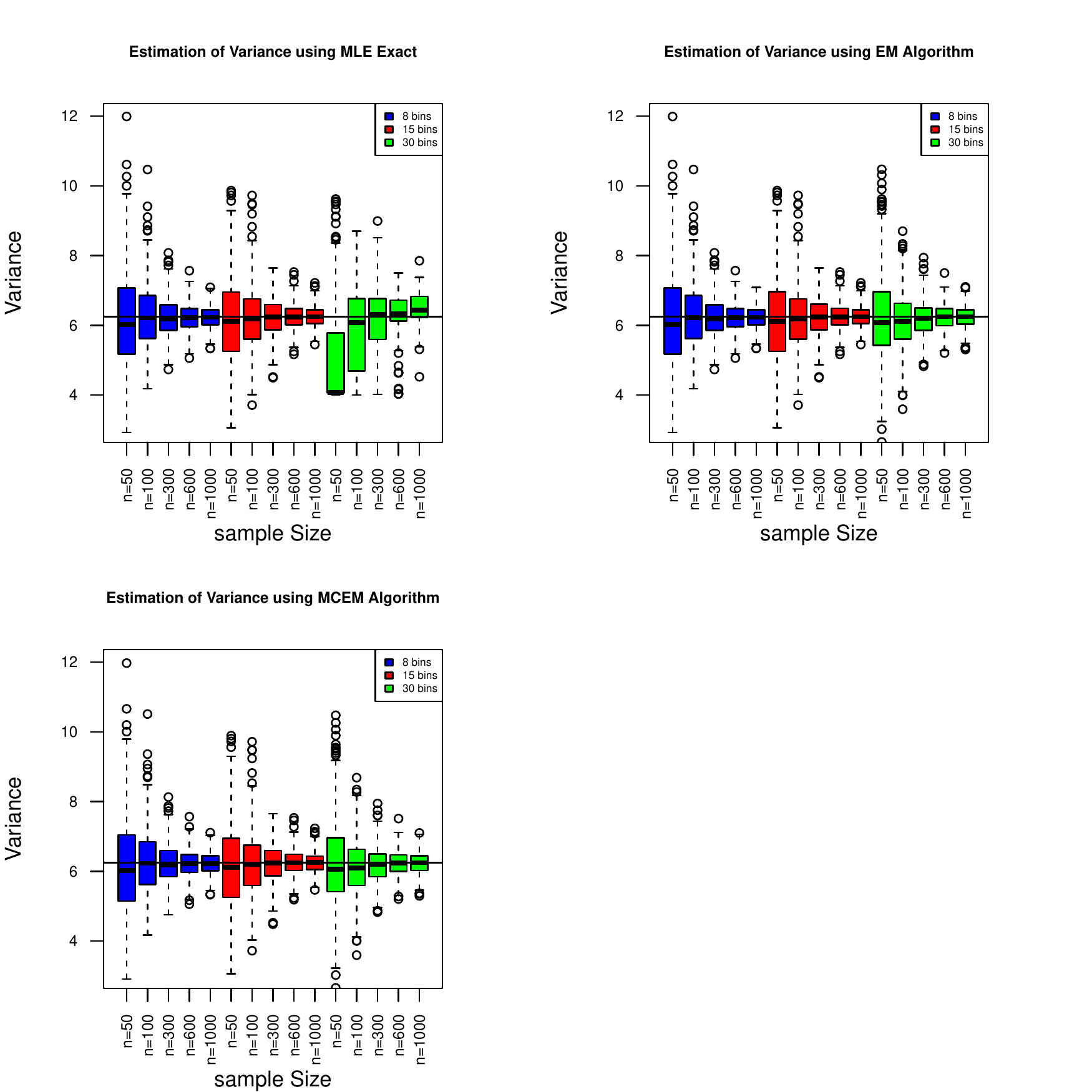}
\caption{\textit{Simulation results: univariate case.} Variance estimates for $k=8$, 15 and 30 intervals (bins) for sample sizes $n=50,100,300,600,1000$. True variance value $\sigma^2=6.25$.}\label{fig:Variance}
\end{figure}

The parameters ($\mu$ and $\sigma$) are estimated using the three methods described in Section 2.1: Exact MLE, EM algorithm and MCEM algorithm. For all the methods, we set the initial values of the parameters as $\mu=67,\sigma=2$.
The root mean squared error (RMSE) of $\mu$ and $\sigma$ over 500 samples are presented in Tables \ref{univariate_Means} and \ref{univariate variances}. Box plots of the parameter estimates obtained across all different scenarios are shown in Figures \ref{fig:MU} and \ref{fig:Variance}. We can observe that for all parameters and all bin sizes the RMSE of the estimates of Exact MLE, EM, and MCEM decrease as the sample size $n$ increases. 

To evaluate the performance of our proposed standard errors for the mean estimates, we calculate the empirical coverage (EC) of 95\% confidence intervals of the form $\hat{\mu}\pm 1.96 se(\hat{\mu})$. We observe in Table \ref{STDmean_est} that most of the ECs are close to the established confidence level of 95\%. In addition, we can observed that the standard deviations of the mean estimates are close to the mean of the proposed standard errors as expected. 

\subsubsection{Bivariate Simulation}
For bivariate data we simulated 500 datasets for each sample size $n$ of 50, 100, 300, 600, and 1000 with 10 equal intervals for each variable ($X_1$ and $X_2$) resulting in 100 rectangles and $2500$ datasets. Datasets are simulated from a bivariate normal with parameters $\mu=(68,68)$ and $\Sigma=\begin{bmatrix} 3& 2 \\ 2 & 6  \end{bmatrix}$. The initial values selected for exact MLE, EM and MCEM methods are $\mu=(67,67)$  and $\Sigma=\begin{bmatrix} 3.2& 2.227106 \\ 2.227106 & 6.2  \end{bmatrix}$. According to \cite{Booth1999} and \cite{McCulloch1997}, the number of Monte-Carlo simulations to obtain the MCEM estimates was fixed to $m=5000$ as the number of MCEM iterations for bivariate data was more than 40. 

\begin{figure}[ht]
\centering
\includegraphics[width=1.15\textwidth]{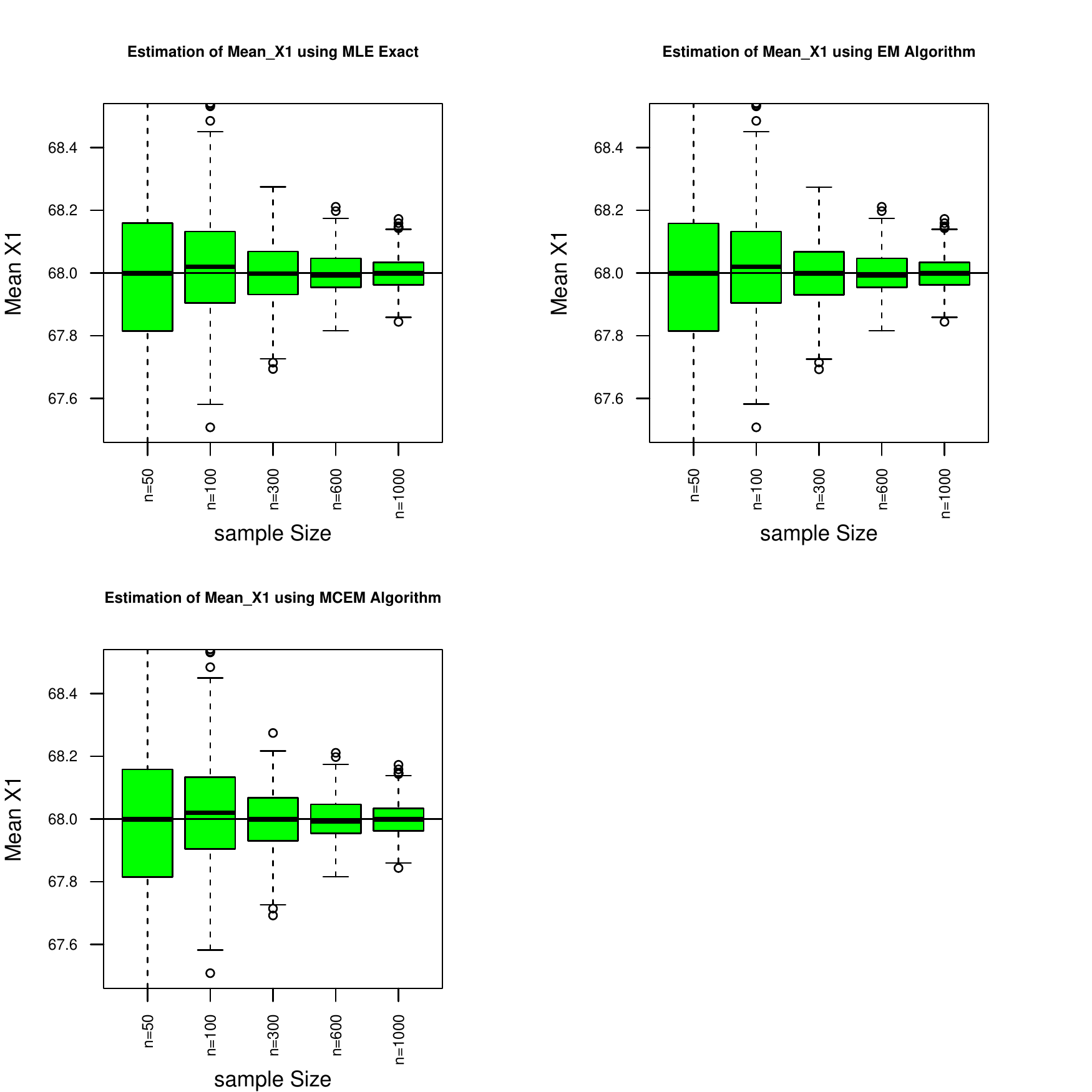}
\caption{\textit{Simulation results: bivariate case.} Estimates of $\mu_{x_1}$ for sample sizes of $n=50,100,300,600,1000$, and $k=10$ intervals for each variable. The horizontal solid line corresponds to the true value $\mu_{x_1}=68$.}
\label{fig:Bivariate-MX1}
\end{figure}

\begin{figure}[ht]
\centering
\includegraphics[width=1.15\textwidth]{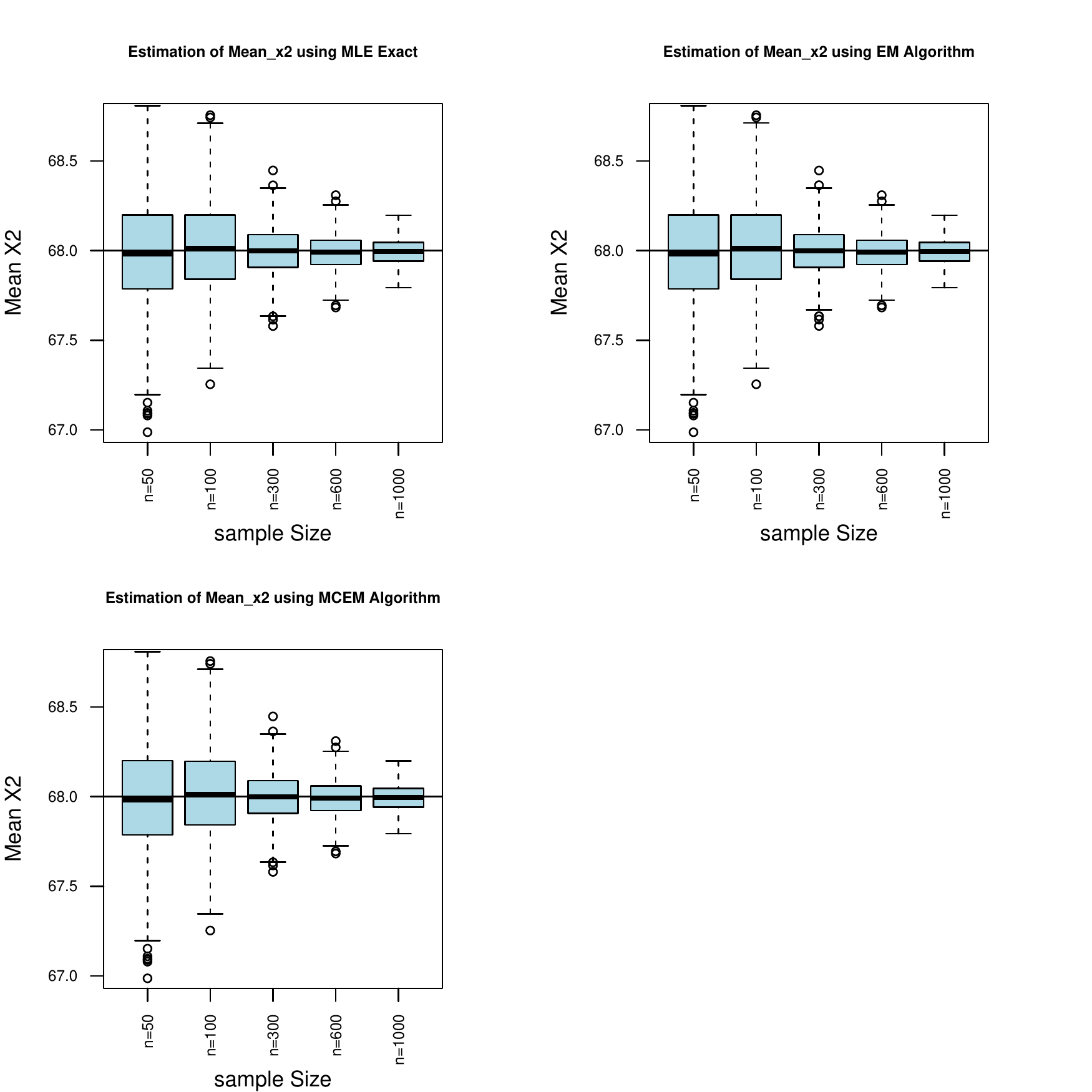}
\caption{\textit{Simulation results: bivariate case.} Estimates of $\mu_{x_2}$ for sample sizes of $n=50,100,300,600,1000$, and $k=10$ intervals for each variable. The horizontal solid line corresponds to the true value $\mu_{x_2}=68$.}
\label{fig:Bivariate-MX2}
\end{figure}

\begin{figure}[ht]
\label{fig:Bivariate-VX1}
\centering
\includegraphics[width=1.15\textwidth]{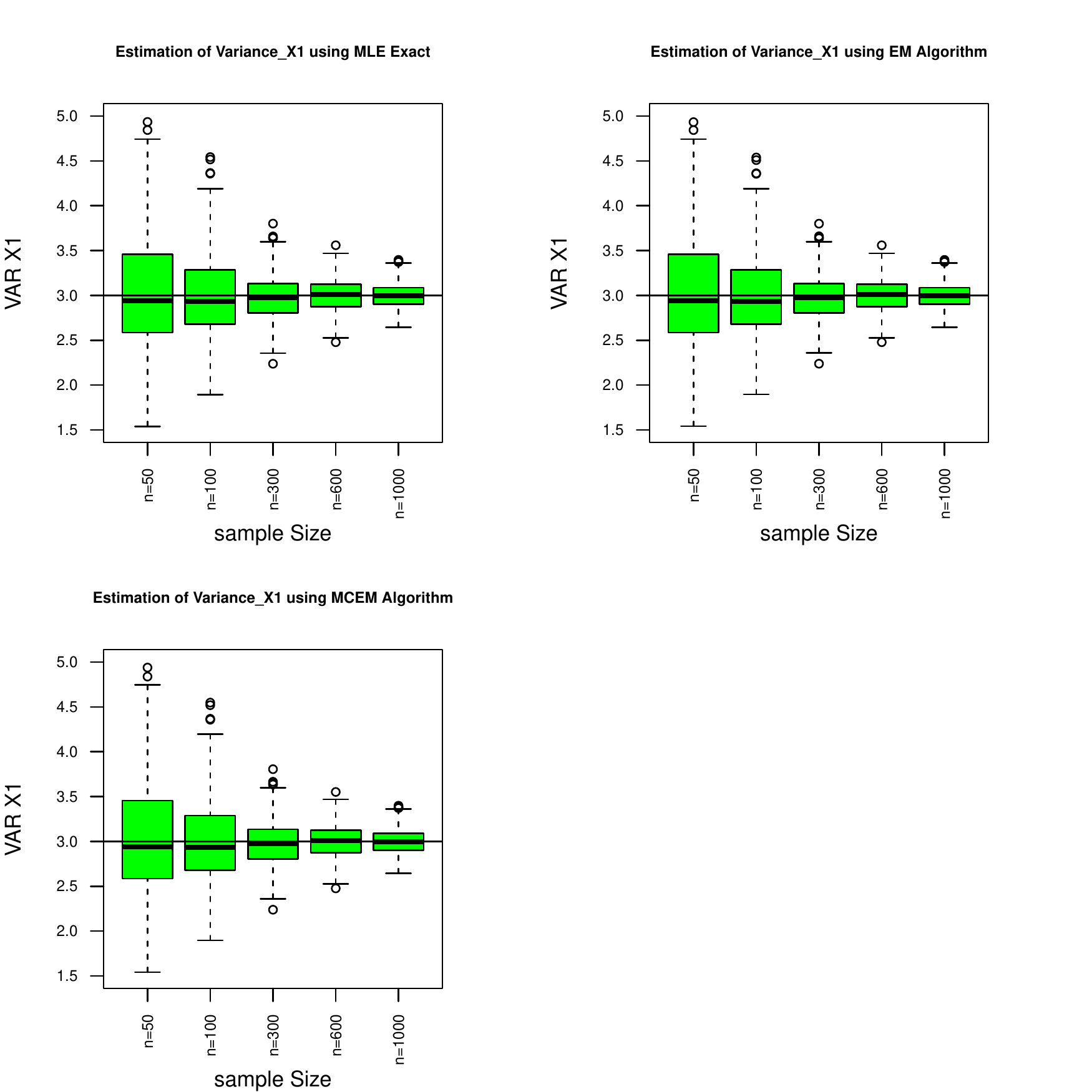}
\caption{\textit{Simulation results: bivariate case.} Estimates of $\sigma_{x_1}^2$ for sample sizes of $n=50,100,300,600,1000$, and $k=10$ intervals for each variable. The horizontal solid line corresponds to the true value $\sigma^2_{x_1}=3$.}
\end{figure}

\begin{figure}[ht]
\label{fig:Bivariate-VX2}
\centering
\includegraphics[width=1\textwidth]{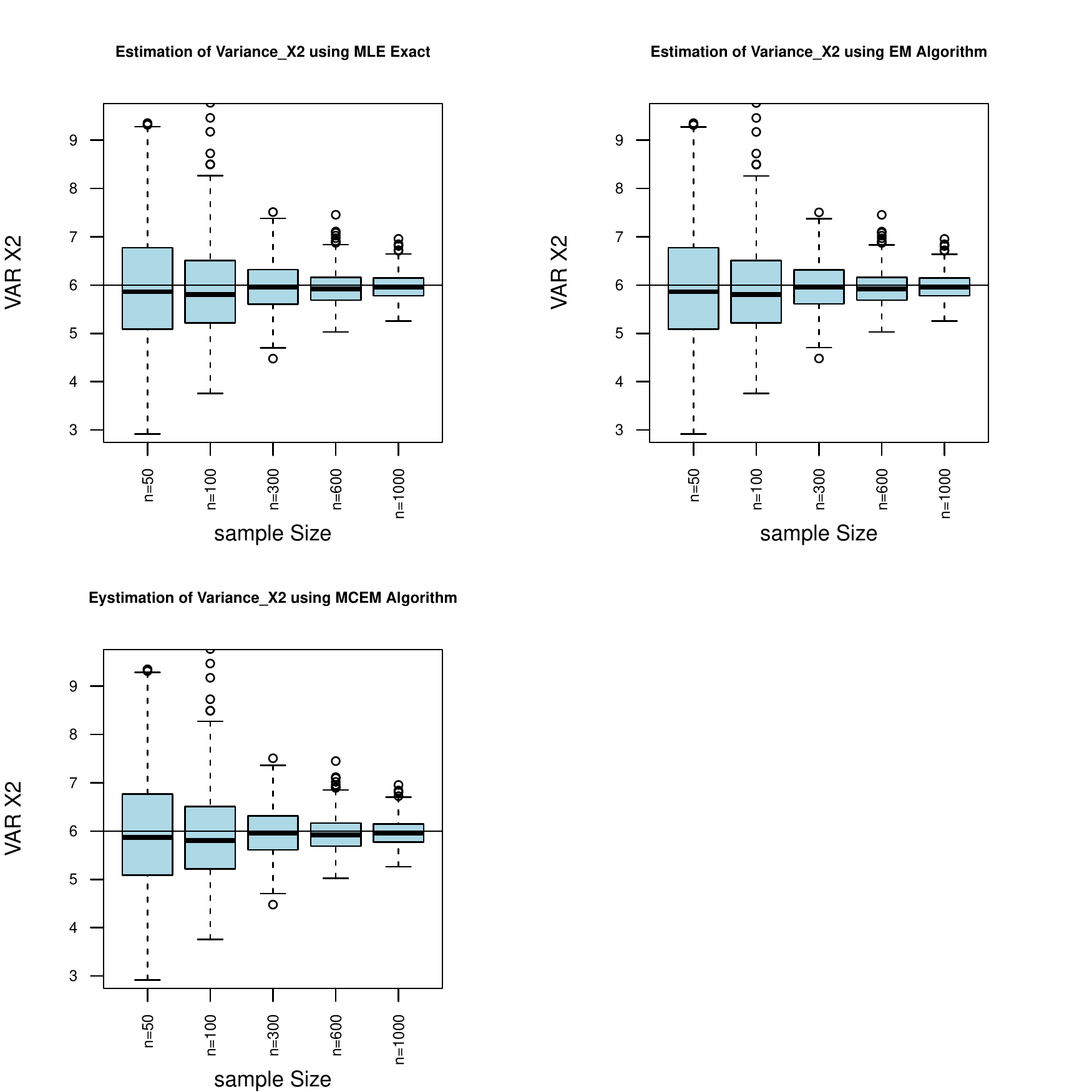}
\caption{\textit{Simulation results: bivariate case.} Estimates of $\sigma_{x_2}^2$ for sample sizes of $n=50,100,300,600,1000$, and $k=10$ intervals for each variable. The horizontal solid line corresponds to the true value $\sigma^2_{x_2}=6$.}
\end{figure}

\begin{figure}[ht]
\centering
\includegraphics[width=1.15\textwidth]{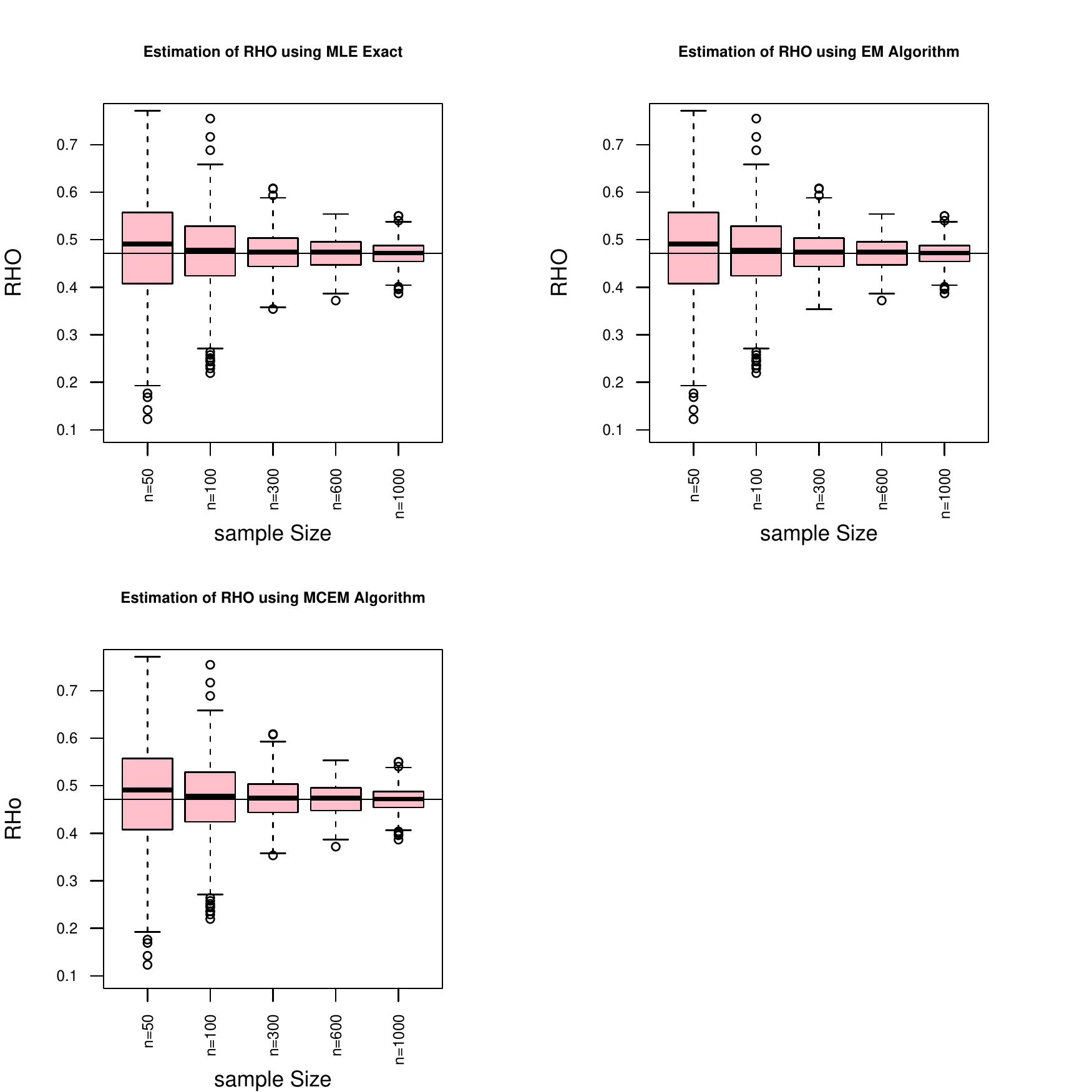}
\caption{\textit{Simulation results: bivariate case.} Estimates of $\rho$ for sample sizes of $n=50,100,300,600,1000$, and $k=10$ intervals for each variable. The horizontal solid line corresponds to the true value of $\rho_{x_1,x_2}$.}
\label{fig:Bivariate-Rho}
\end{figure}

Figures \ref{fig:Bivariate-MX1} to \ref{fig:Bivariate-Rho} present the box plots of the parameter estimates for each method and different sample sizes.  Our results also show that the Exact MLE, EM and MCEM yielded very similar estimates as expected even for the smaller $n$ of 50. In addition, we can observe in Table \ref{Results_Biv} that the root mean squared error (RMSE) of the estimates decrease as the sample size $n$ increases for all parameters and methods. Table \ref{STDMean_X1} shows the ECs for 95\% confidence intervals of the form $\hat{\mu}\pm 1.96 se(\hat{\mu})$ for both $\mu_{x_1}$ and $\mu_{s_2}$. We observe that in most cases the ECs are close to the established 95\% level of confidence for both EM and MCEM methods.

\section{Discussion on Conclusion}
We have proposed three approaches, namely, Exact MLE, EM and MCEM algorithms to estimate the parameters of normally distributed grouped data. The cases of univariate, bivariate and multivariate normal were considered and parameter estimates using each method were presented. For the exact MLE approach, by considering the distribution of the counts to be multinomial, with probabilities based on the normal CDFs, the exact data log-likelihood could be formulated and the MLE values could be found using numerical methods. 
For EM and MCEM algorithms, using the exact observed data log-likelihood, the complete data log-likelihood was computed and the parameter estimates obtained in closed forms using the formulas in Sections 2.1.2, 2.1.3, 2.2.2, 2.2.3 and 2.2.4. 

To compare the methods, first, we considered the well-known Galton data, and parameter estimates were found for the cases of univariate and bivariate grouped data. Next, the mean absolute relative differences between the estimates obtained by Exact MLE and each of the other methods (EM and MCEM) were calculated and showed that EM led to the closest results to the exact MLE. Then, simulation studies were implemented for the univariate and bivariate cases for different scenarios. For most parameters, the results from the EM and MCEM algorithms were similar to the ones from the exact MLE as expected by their convergence properties shown in Chapters 1 and 3 of \cite{Mclachlan2007}.

Based on our results, we conclude that there are some advantages and drawbacks regarding the three methods. The exact MLE method leads to efficient and unbiased estimates; however, there is no closed-form for the parameter estimates, and they are found using numerical optimization methods. Moreover, this method is highly sensitive to the optimization method and initial values. In comparison, the EM and MCEM methods in our analyses were not as sensitive to initial values as the Exact MLE method. In addition, for both EM and MCEM algorithms, there are specific and closed formulae for the parameter estimates. 




\section*{Acknowledgments}

This work was supported by the National Science and Engineering Research Council of Canada.

\bibliography{references}

\clearpage

\begin{table}[ht]
\centering
\caption{\small{\textit{Simulation results: univariate case.}  RMSE of mean estimates of 500 simulated samples for $n=50,100,300,600,$ and $1000$ and number of intervals (bins) $k = 8, 15,$ and 30 over three estimation methods.}}
\begin{tabular}{|lllll|}
  \hline
   & & \multicolumn{3}{c|}{RMSE for Means}   \\ 
  
  Method & $n$ & $k=8$ & $k=15$ & $k=30$ \\ 
  \hline 
 Exact MLE & 50 & 0.34368 & 0.34848 & 0.80577 \\ 
  & 100 & 0.25917 & 0.25484 & 0.73227 \\ 
  & 300 & 0.13859 & 0.15677 & 0.62772 \\ 
   & 600 & 0.10698 & 0.10678 & 0.33962 \\ 
  & 1000 & 0.07972 & 0.08207 & 0.22194 \\ \hline 
  EM & 50 & 0.34369 & 0.34849 & 0.37453 \\ 
  & 100 & 0.25917 & 0.25485 & 0.25459 \\ 
  & 300 & 0.13859 & 0.15678 & 0.14536 \\ 
  & 600 & 0.10697 & 0.10678 & 0.10202 \\ 
  & 1000 & 0.07972 & 0.08207 & 0.07917 \\ \hline
  MCEM & 50 & 0.34369 & 0.34845 & 0.37479 \\ 
  & 100 & 0.25922 & 0.25501 & 0.25457 \\ 
  & 300 & 0.13893 & 0.15668 & 0.14526 \\ 
  & 600 & 0.10705 & 0.10687 & 0.1022 \\ 
   & 1000 & 0.07976 & 0.08243 & 0.07925 \\ 
   \hline
\end{tabular}
\label{univariate_Means} 
\end{table}

\newpage
\begin{table}[ht]
\centering
\small
\caption{\small{\textit{Simulation results: univariate case.} Average standard error (SE) and empirical coverage (EC) (over 500 simulated datasets) for the EM and MCEM estimates of $\mu$ for $n=50,100,300,600,1000$, and $k=15$ number of intervals (bins).}}
\begin{tabular}{|lllll|}
  \hline
   & & \multicolumn{3}{c|}{Standard Errors for Mean Estimates}   \\ 
  
 $n$ & Method & Ave. $\hat{\mu}$ (std $\hat{\mu}$) &  Ave. SE for $\hat{\mu}$  & EC \\ 
  \hline
50 & EM & 68.00287872  (0.34882416) & 0.36509683 & 94.8 \\ 
  & MCEM & 68.00295812  (0.34879046) & 0.35490471  & 94.6 \\ \hline 
   100 & EM & 67.99932895  (0.25510392) & 0.25597085  & 94.4 \\ 
   & MCEM & 67.99935779  (0.25526480) & 0.25237501  & 93.8 \\ \hline
   300 & EM & 68.00202742  (0.15692531) & 0.14679516  & 92.8 \\ 
   & MCEM & 68.00153299  ( 0.15683075) & 0.14610848  & 92.4 \\ \hline
   600 & EM & 67.99406868  (0.10672629) & 0.10381674  & 94.2 \\ 
   & MCEM & 67.99370576   (0.10679113) & 0.10354026  & 94 \\\hline 
   1000 & EM & 67.99913644  (0.08215109) & 0.080350265  & 93.8 \\ 
    & MCEM & 67.99914431  (0.08250915) & 0.080233290  & 93.8 \\ \hline
\end{tabular}
\label{STDmean_est}
\end{table}

\newpage

\begin{table}[ht]
\centering
\small
\caption{\small{\textit{Simulation results: univariate case.} RMSE of variance estimates of 500 simulated samples for $n=50,100,300,600,$ and $1000$ and number of intervals (bins) $k = 8, 15,$ and 30 over three estimation methods.}}
\begin{tabular}{|lllll|}
 \hline
    & & \multicolumn{3}{c|}{RMSE for Variances}   \\ 
  
 Method & $n$ & $k=8$ & $k=15$ & $k=30$ \\ 
  \hline
 Exact MLE & 50 & 1.40998 & 1.26072 & 1.90078 \\ 
    & 100 & 0.93567 & 0.89101 & 1.2376 \\ 
   & 300 & 0.54813 & 0.50938 & 1.00808 \\ 
   & 600 & 0.39345 & 0.35808 & 0.49013 \\ 
   & 1000 & 0.30994 & 0.29212 & 0.45055 \\  \hline
  EM & 50 & 1.40998 & 1.2607 & 1.28548 \\ 
   & 100 & 0.93576 & 0.89092 & 0.86004 \\ 
   & 300 & 0.54812 & 0.50927 & 0.5132 \\ 
   & 600 & 0.39336 & 0.35814 & 0.35481 \\ 
  & 1000 & 0.31006 & 0.29215 & 0.29758 \\ \hline
  MCEM & 50 & 1.40989 & 1.26103 & 1.2848 \\ 
   & 100 & 0.93697 & 0.89189 & 0.86105 \\ 
  & 300 & 0.55068 & 0.50973 & 0.51382 \\ 
   & 600 & 0.39327 & 0.35742 & 0.3557 \\ 
   & 1000 & 0.31167 & 0.29247 & 0.29938 \\ \hline
\end{tabular}
\label{univariate variances}
\end{table}




\newpage
\begin{table}[ht]
\centering
\small
\caption{\small {Root mean squared errors (RMSE) of bivariate parameters ($\mu_{x_1}$,$\mu_{x_2}$,$\sigma^2_{x_1}$,$\sigma^2_{x_2}$,$\rho$) across 500 data sets for each sample size $n=50,100,300,600,1000$ with 10 intervals for each variable (100 rectangles) and three methods used.}} 
\begin{tabular}{|lllll|}
  \hline
  parameter & sample size & Exact MLE & EM & MCEM \\ 
  \hline 
$\mu_{x_1}$ & 50 & 0.252397 & 0.252381 & 0.252402 \\  
    & 100 & 0.17686 & 0.176857 & 0.176825 \\ 
    & 300 & 0.099135 & 0.099115 & 0.099109 \\ 
    & 600 & 0.06756 & 0.067556 & 0.067578 \\ 
    & 1000 & 0.054078 & 0.054073 & 0.054092 \\ \hline 
   $\mu_{x_2}$ & 50 & 0.337363 & 0.337353 & 0.337348 \\ 
   & 100 & 0.250034 & 0.250038 & 0.250025 \\ 
    & 300 & 0.140965 & 0.140715 & 0.140641 \\ 
    & 600 & 0.101405 & 0.101408 & 0.101358 \\ 
    & 1000 & 0.075395 & 0.075394 & 0.075369 \\ \hline
   $\sigma^2_{x_1}$ & 50 & 0.636113 & 0.635723 & 0.63575 \\ 
    & 100 & 0.438604 & 0.438275 & 0.438731 \\ 
    & 300 & 0.244176 & 0.244165 & 0.244096 \\ 
    & 600 & 0.187619 & 0.187514 & 0.187396 \\ 
    & 1000 & 0.138219 & 0.13813 & 0.138106 \\ \hline
   $\sigma^2_{x_2}$& 50 & 1.306861 & 1.305404 & 1.30616 \\ 
  & 100 & 0.953621 & 0.952149 & 0.953117 \\ 
    & 300 & 0.519407 & 0.519655 & 0.520149 \\ 
    & 600 & 0.378151 & 0.376675 & 0.376777 \\ 
    & 1000 & 0.287702 & 0.286377 & 0.287136 \\ \hline
   $\rho$ & 50 & 0.115697 & 0.115659 & 0.115676 \\ 
    & 100 & 0.081728 & 0.081703 & 0.081697 \\ 
   & 300 & 0.044546 & 0.044568 & 0.044561 \\ 
    & 600 & 0.033404 & 0.033389 & 0.033383 \\ 
   & 1000 & 0.026724 & 0.026709 & 0.026731 \\ 
   \hline
\end{tabular}

\label{Results_Biv}
\end{table}

\begin{table}[ht]
\centering
\small
\caption {\small{ \textit{Simulation results: bivariate case.} Average standard error (SE) and empirical coverage (EC) (over 500 simulated datasets) for the EM and MCEM estimates of $\mu_{X_1}$ and $\mu_{x_2}$ for $n=50,100,300,600,1000$, and 100 rectangles.}}
\begin{tabular}{ |llllll|}
  \hline
  Parameter & $n$ & method & Ave. $\hat{\mu}$ (sd $\hat{\mu}$)& Ave. SE for $\hat{\mu}$ & EC \\ 
  \hline \hline
$\mu_{x_1}$ &50 & EM &67.99029176 (0.25244727) & 0.24601741 & 0.942 \\ 
  & & MCEM & 67.99024615 (0.25246634) & 0.23884756  & 0.934 \\ \hline
   &100 & EM & 68.01838946 (0.17607445)  & 0.17393201  & 0.936 \\ 
    && MCEM & 68.01840550 (0.17604050)& 0.16872550  & 0.934 \\ \hline
   &300 & EM & 67.99715989 (0.09917395)  & 0.10089726  & 0.966 \\ 
    && MCEM & 67.99715451 (0.09916721) & 0.09782386  & 0.958 \\ \hline
   &600 & EM & 67.99817344 (0.06759916) & 0.07171556  & 0.966 \\ 
   && MCEM & 67.99813458 (0.06762007) & 0.06942490 & 0.96 \\ \hline
   &1000 & EM & 67.99940268 (0.05412395)  & 0.05552854 & 0.946 \\ 
 & &MCEM & 67.99927880 (0.05414088) & 0.0537654701  & 0.942 \\ \hline 
   \hline
   $\mu_{x_2}$&50 & EM & 67.98844687 (0.33749244) & 0.34712279  & 0.934 \\ 
    && MCEM & 67.98851100(0.33749043)& 0.28934334  & 0.892 \\ \hline 
   &100 & EM & 68.01415707 (0.24988641) & 0.24504268 & 0.95 \\ 
&    & MCEM & 68.01411880 (0.24987648)  & 0.20503459  & 0.898 \\ \hline 
 &  300 & EM & 67.99908176 (0.14085275) & 0.14285627  & 0.958 \\ 
  & & MCEM & 67.99912894 (0.14077926)  & 0.11950836  & 0.9 \\ \hline
   &600 & EM & 67.99156129 (0.10115777)  & 0.10078140 & 0.948 \\ 
    && MCEM & 67.99151371 (0.10110339)  & 0.08452820  & 0.892 \\ \hline 
   &1000 & EM & 67.99395405 (0.07522677) & 0.07822683   &0.954 \\ 
   && MCEM & 67.99405000 (0.07520861)  & 0.06556686  & 0.918 \\ \hline

\end{tabular}

\label{STDMean_X1}
\end{table}

\newpage
\newpage

\newpage

\newpage

\newpage

\newpage

\clearpage
\begin{appendix}
\section{Expectations for the E-step of the EM algorithm for univariate grouped data}
\label{Sec:Univariate}
In Section 2.1.2, to find the updated estimates of the parameters, we have to calculate the following expectations w.r.t the density  $\frac{f(x;\theta)}{P_i(\theta)}$:
$$E_{\theta^{(p)}}\Big(X|X\in \mathcal{X}_i\Big)$$
and
$$E_{\theta^{(p)}}\Big((X-\mu^{(p+1)})^2|X\in\mathcal{X}_i\Big),$$
where $f(x;\theta)$ is the univariate normal distribution.
Let $\mathcal{X}_i=(a,b)$, these expectations can be obtained as follows:
\begin{eqnarray*}
E_{\theta^{(p)}}\Big [X|X\in\mathcal{X}_i\Big ]=[F(b)-F(a)].E(X)&=&\int \limits_{a}^{b}x \frac{1}{\sqrt{2\pi}\sigma^{(p)}}e^{-\frac{1}{2\sigma^{2(p)}}(x-\mu^{(p)})^2}dx\\
&=&\int \limits_{a^*}^{b^*}(\sigma^{(p)} t+\mu^{(p)}) \frac{1}{\sqrt{2\pi}}e^{-\frac{1}{2}t^2}dt\\
&=&\sigma^{(p)} \int\limits_{a^*}^{b^*}t \frac{1}{\sqrt{2\pi}}e^{-\frac{1}{2}t^2}dt+\mu^{(p)} \int\limits_{a^*}^{b^*} \frac{1}{\sqrt{2\pi}}e^{-\frac{1}{2}t^2}dt \\
&=& \mu^{(p)}\Big[F(b^*)-F(a^*) \Big]-\sigma^{(p)}\Big[f(b^*)-f(a^*)\Big],
\end{eqnarray*}
\noindent where
$t=\frac{x-\mu^{(p)}}{\sigma^{(p)}}$,
$a^*=\frac{a-\mu^{(p)}}{\sigma^{(p)}}$, and $b^*=\frac{b-\mu^{(p)}}{\sigma^{(p)}}$. See also \cite{AS254}.

\begin{eqnarray*}
[F(b)-F(a)]E(X-\mu^{(p+1)})^2&=&\int \limits_{a}^{b}(x-\mu^{(p+1)})^2 \frac{1}{\sqrt{2\pi}\sigma^{(p)}}e^{-\frac{1}{2\sigma^{2(p)}}(x-\mu^{(p)})^2}dx=\\
&&\int\limits_{a}^{b}x^2 \frac{1}{\sqrt{2\pi}\sigma^{(p)}}e^{-\frac{1}{2\sigma^{2(p)}}(x-\mu^{(p)})^2}dx-\\
&&2\mu^{(p+1)}\int\limits_{a}^{b}x \frac{1}{\sqrt{2\pi}\sigma^{(p)}}e^{-\frac{1}{2\sigma^{2(p)}}(x-\mu^{(p)})^2}dx.\\
&&\mu^{2(p+1)}\int \limits_{a}^{b} \frac{1}{\sqrt{2\pi}\sigma^{(p)}}e^{-\frac{1}{2\sigma^{2(p)}}(x-\mu^{(p)})^2}dx\mbox{.}
\end{eqnarray*}


Now let $a^*=\frac{a-\mu^{(p)}}{\sigma^{(p)}}$,  $b^*=\frac{b-\mu^{(p)}}{\sigma^{(p)}}$, $t=\frac{x-\mu^{(p)}}{\sigma^{(p)}}$, and using
\begin{eqnarray*}
&&\int \limits_{a}^{b}x^2 \frac{1}{\sqrt{2\pi}\sigma^{(p)}}e^{-\frac{1}{2\sigma^{2(p)}}(x-\mu^{(p)})^2}dx=\\
&&\int \limits_{a^*}^{b^*}(\sigma^{(p)} t+\mu^{(p)})^2 \frac{1}{\sqrt{2\pi}}e^{-\frac{1}{2}t^2}dt=\\
&&\sigma^{2(p)} \int\limits_{a^*}^{b^*}t^2 \frac{1}{\sqrt{2\pi}}e^{-\frac{1}{2}t^2}dt+\mu^{2(p)} \int\limits_{a^*}^{b^*} \frac{1}{\sqrt{2\pi}}e^{-\frac{1}{2}t^2}dt+\\
&&2\sigma^{(p)}\mu^{(p)}\int\limits_{a^*}^{b^*}t \frac{1}{\sqrt{2\pi}}e^{-\frac{1}{2}t^2}dt\\
&&\sigma^{2(p)}\Bigg[[F(b^*)-F(a^*)]-[b^*f(b^*)-a^*f(a^*)]\Bigg]+\\
&&\mu^{2(p)}[F(b^*)-F(a^*)]-2\sigma^{(p)}\mu^{(p)}[f(b^*)-f(a^*)],
\end{eqnarray*}
\noindent we obtain
\begin{eqnarray*}
E(X-\mu^{(p+1)})^2&&=\frac{1}{F(b^*)-F(a^*)}\times \\
&&\sigma^{2(p)}\Bigg[\Big(F(b^*)-F(a^*)\Big)-\Big(b^*f(b^*)-a^*f(a^*)\Big) \Bigg]+\\ &&\Big(\mu^{(p+1)}-\mu^{(p)}\Big)^2\Big[F(b^*)-F(a^*)\Big]+\\
&&2\sigma^{(p)}\Bigg[\Big(\mu^{(p+1)}-\mu^{(p)}\Big)\Big[f(b^*)-f(a^*)\Big]\Bigg].
\end{eqnarray*}

\section{Finding $E_{\Theta^{(p)}}\Big[X_i\Big]$ and $E_{\Theta^{(p)}}\Big[(X_iX_j^T)\Big]$ for Multivariate Normal}
\label{Sect::Multivariate}

In Sections 2.2.2 and 2.2.4, to find the exact estimate of the parameters using the EM approach, the expectations in the general form of 
$$E_{\Theta^{(p)}}\Big[X_i \Big|(X_1,\dots,X_d)\in (\mathcal{X}_1,\dots \mathcal{X}_d) \Big]$$
and
$$E_{\Theta^{(p)}}\Big[(X_i-\mu^{(p+1)})(X_j-\mu^{(p+1)})^T \Big|(X_1,\dots,X_d)\in (\mathcal{X}_1,\dots \mathcal{X}_d) \Big]$$
should be found. In what  follows we present the main steps of the calculations. For further details see  \cite{Manjunath2021}.

Using the moment generation function for the multivariate normal, the expectations can be obtained as follows: 
\begin{equation}\label{EQ23}
E(X_i)=\frac{\partial m(t)}{\partial t_i}|_{t=0}=\sum_{k=1}^{d} \sigma_{i,k} (F_k(a_k)-F_k(b_k))
\end{equation}
where
\begin{eqnarray*}
&&F_i(x)=\\
&&\int \limits_{a_1^*}^{b_1^*} \dots \int \limits_{a_{i-1}^*}^{b_{i-1}^*} \int \limits_{a_{i+1}^*}^{b_{i+1}^*} \dots \int \limits_{a_d^*}^{b_d^*} \phi_{\alpha \Sigma}(x_1,\dots,x_{i-1},x,x_{i+1},\dots,x_d) dx_d\dots dx_{i+1} dx_{i-1} \dots dx_1
\end{eqnarray*}
$$a_i^*=a_i-\sum_{k=1}^{d} \sigma_{i,k} t_k , $$
$$b_i^*=b_i-\sum_{k=1}^{d} \sigma_{i,k} t_k  $$
and at $t_k=0$, for all $k=1,2,\dots,d$, $a_i^*=a_i$ and $b_i^*=b_i$. 
It should be noted that in $F_i(x)$, $\phi_{\alpha, \Sigma}(x)$ is coming from:
\[\phi_{\alpha,\mu,\Sigma}(x)  = \left\{ \begin{array}{ll}
\frac{\phi_{\mu,\Sigma}(x)}{P(a\leq X \leq b)} & \mbox{for $a \leq x \leq b$}\\
0 & \mbox{otherwise}\end{array} \right. \]
$$m(t)=e^T \Phi_{\alpha \Sigma}$$
for $\zeta=\Sigma t$:
$$\Phi_{\alpha \Sigma}=\frac{1}{\alpha (2\pi)^{(\frac{d}{2})}|\Sigma|^{\frac{1}{2}}} \int \limits_{a-\zeta}^{b-\zeta}exp\Big(-\frac{1}{2}x^T \Sigma^{-1}x\Big)dx $$

\noindent for the case of $\mu=0$ and $\alpha=P(a<X<b)$.
Considering $Y\sim N(\mu,\Sigma)$ with $a^*<y<b^*$, then using the transformation, $X=Y-\mu \sim N(0,\Sigma)$ which change within the range of $a=a^*-\mu<x<b^*-\mu=b$. So, for the general case $\mu$ (not the case of $\mu=0$), using the transformation idea for the expectation, we will have $E(Y)=E(X)+\mu$, then for the multivariate normal expectation we will obtain:
$$E(Y_i)=\sum_{k=1}^{d} \sigma_{i,k} (F_k(a_k)-F_k(b_k))+\mu_i$$
Similarly, we can show that for all $t_k=0,\,k=0,\dots,d$ (see \cite{Manjunath2021}) we obtain:
\begin{eqnarray*}
E(X_i X_j)&=& \frac{\partial ^2 m(t)}{\partial t_j \partial t_i}|_{t=0}=\sigma_{i,j}+\sum_{k=1}^{d} \sigma_{i,k} \frac{\sigma_{j,k}(a_kF_k(a_k)-b_kF_k(b_k))}{\sigma_{k,k}}\\
&&+\sum_{k=1}^{d}\sigma_{i,k} \sum_{q \neq k}
(\sigma_{j,q}-\frac{\sigma_{k,q} \sigma_{j,k}}{\sigma_{k,k}})\Bigg[\Big(F_{k,q}(a_k,a_q)-F_{k,q}(a_k,b_q)\Big)\\
&&-\Big(F_{k,q}(b_k,a_q)-F_{k,q}(b_k,b_q)\Big)\Bigg]    
\end{eqnarray*}
where 
\begin{eqnarray*}
&&F_{k,q}(x,y)=\\
&&\int \limits_{a_1^*}^{b_1^*} \dots \int \limits_{a_{k-1}^*}^{b_{k-1}^*} \int \limits_{a_{k+1}^*}^{b_{k+1}^*} \dots
\int \limits_{a_{q-1}^*}^{b_{q-1}^*} \int \limits_{a_{q+1}^*}^{b_{q+1}^*} \dots
\int \limits_{a_d^*}^{b_d^*} \phi_{\alpha \Sigma}(x,y,x_{-k,-q})dx_{-k,-q^\prime}
\end{eqnarray*}
and 
$$x_{-k,-q}=(x_1,\dots,x_{k-1},x_{k+1},\dots,x_{q-1},x_{q+1},\dots,x_d)^\prime \mbox{} $$
for $k \neq q \mbox{.}$
As the covariance matrix is invariant to the shift of the variables we will have
$$cov(Y_i,Y_j)=cov(X_i,X_j)=E(X_i X_j)-E(X_i)E(X_j)$$
All of these expectations are calculated at the current state of the parameters $\mu^{(p)}$ and $\Sigma^{(p)}$.

\end{appendix}

\end{document}